# Characteristics of grassy ELMs and its impact on the divertor heat flux width

Nami Li[1,2], X.Q. Xu[2], N. Yan[3], Y.F. Wang[3], J.Y. Zhang[3], J.P. Qian[3], J.Z. Sun[2] and D.Z. Wang[2]

[1]*Lawrence Livermore National Laboratory, Livermore, CA 94550, USA*
[2]*School of Physics, Dalian University of Technology, Dalian 116024, China*
[3]*Institute of Plasma Physics, Chinese Academy of Sciences, Hefei 230031, China*

**Abstract**

BOUT++ turbulence simulations are conducted for a 60s steady-state long pulse high $\beta_p$ EAST grassy ELM discharge. BOUT++ linear simulations show that the unstable mode spectrum covers a range of toroidal mode numbers from low-n (n=10~15) peeling-ballooning modes (P-B) to high-n (n=40~80) drift-Alfvén instabilities. Nonlinear simulations show that the ELM crash is trigged by low-n peeling modes and fluctuation is generated at the peak pressure gradient position and radially spread outward into the Scrape-Off-Layer (SOL), even though the drift-Alfvén instabilities dominate the linear growth phase. However, drift-Alfvén turbulence delays the onset of the grassy ELM and enhances the energy loss with the fluctuation extending to pedestal top region. Simulations further show that if the peeling drive is removed, the fluctuation amplitude drops by an order of magnitude and the ELM crashes disappear. The divertor heat flux width is ~2 times larger than the estimates based on the HD model and the Eich's ITPA multi-tokamak scaling (or empirical Eich scaling) due to the strong radial turbulence transport.

## 1. Introduction

Simultaneous control of large Edge-Localized-Modes (ELMs) and high divertor heat load in H-mode plasma is crucial for steady-state operation of a tokamak fusion reactor[1]. The grassy ELMs, one of small ELMs, are characterized by a high frequency and spatially localized quasi-periodic collapse in the bottom of pedestal near the separatrix. The peak heat fluxes on the divertor target plate for grassy ELMs are reduced by more than 90% as compared to type-I ELMs[2]. Recent experiments, from AUG, DIII-D, EAST, and TCV, show that small ELM regimes have the following desired features: (1) good energy confinement; (2) divertor heat flux comparable with Type-I inter-ELM level; (3) significantly broadening of divertor heat flux profile; (4) quasi-continuous particle and power exhaust to the divertors. H-mode plasma regimes with small ELMs compatible with high density, high confinement, and power exhaust compatible H-mode regimes with small ELMs have been achieved in TCV and ASDEX Upgrade (AUG)[3]. The Quasi-Continuous Exhaust (QCE) regime in AUG is observed with enhanced filamentary transport due to small type-II ELMs and significantly broadened power fall-off length $\lambda_q$ (by a factor up to 4), by operation with increasing fueling[4-6]. JOREK simulations have successfully reproduced type-I ELM cycles and the QCE regime in AUG[7]. Recent DIII-D grassy ELM experiments show a consistent divertor heat flux width broadening and peak amplitude reduction. From the inter-ELM phase to the grassy ELM phase with Resonant Magnetic Perturbations (RMP) 3D fields, the width on the inner divertor target increases about 2 times with low RMP current $I_{RMP}$=1950kA and about 4 times with high $I_{RMP}$=2450kA[8,9]. The grassy-ELMs exhibit a reduced peak heat flux to the divertor similar to the inter-ELM heat flux with good confinement. Since 2016, a high confinement performance grassy-ELM H-mode regime has been achieved in EAST. Heat fluxes of the grassy ELMs are found to be only $1/20^{th}$ - $1/10^{th}$ of those with large Type-I ELMs and they are comparable with inter-ELM levels[2]. BOUT++ simulations for DIII-D, EAST, ITER and CFETR show consistent trends and demonstrate consistent divertor heat flux width broadening and peak amplitude reduction in the grassy ELM regimes[10-14]. A 60s steady-state long pulse high $\beta p$ EAST grassy ELM discharge has been achieved with high performance in 2019[15]. Therefore, H-mode plasma regimes with small ELMs offers a potential solution to future reactor power exhaust as it features many aspects required for the operation of future fusion reactors and has been proposed as the primary ELM-mitigation solution for the Chinese Fusion Engineering Test Reactor (CFETR).

Even though substantial progresses have made in the access conditions of this small ELM regime for some tokamaks, it is non-trivial experiments for achieving the grassy ELM regime for other tokamaks and a detailed physics understanding on the underlying mechanism are not yet available. Whether the grassy ELM regime could be accessed and its compatibility with stationary operations in future fusion reactors is still uncertain, further investigation is required.

In order to investigate the characteristics of the grassy ELM and its impact on the divertor heat flux width, BOUT++ turbulence simulations are conducted for a 60s steady-state long pulse high $\beta p$ grassy ELM discharge with shot #090949 for EAST. The main parameters for this shot are close to the steady state scenario with Q>5 in CFETR Phase 2 performance. The simulation for understanding the physics of this shot is very important for future development of long pulse steady state scenarios for future tokamak reactors. In this study, pedestal linear stability analysis is performed with both ELITE code and BOUT++ turbulence code, and the nonlinear dynamics and divertor heat flux width are calculated by BOUT++ six-field turbulence code. The paper is organized as follows. Section 2 contains a description of simulation settings. The simulation results of characteristics of grassy ELMs for EAST #090949 shot are shown in Section 3, including (1) linear MHD stability analysis, (2) characterization of ELM nonlinear dynamics, and (3) impact of P-B mode and drift-Alfvén mode on the grassy ELMs. Section 4 shows the divertor heat flux width and a summary of the results is given in Sec. 5.

## 2. Equilibria for the EAST high-$\beta_p$ long-pulse experiment and simulation settings

The plasma profiles used in this work are based on the experimental measurement from EAST high-$\beta_p$ long-pulse discharge with upper single null divertor configuration. The equilibrium is reconstructed using the kinetic EFIT code[16,17] with the constraints of experimentally measured total pressure profile and flux surface averaged toroidal current density profile with bootstrap current $j_{BS}$ dominated in the pedestal region. The simulations include the plasma edge, the SOL region and private flux region. The simulation domain is shown in Fig. 1, which ranges from normalized poloidal flux ψ=0.8 to ψ=1.10, where ψ=1.0 is the magnetic separatrix as shown by the red curve in Fig. 1. The spatial resolution of the grid generated from the equilibrium file (EFIT g-file) is 260 radial grid points and 64 poloidal grid points. In the poloidal direction, there are 4 grid points for each divertor leg region from x-point to the divertor plate and 56 grid points for the main plasma region above the x-point. In the toroidal direction, we only simulate one-fifth of the torus for the nonlinear simulations.

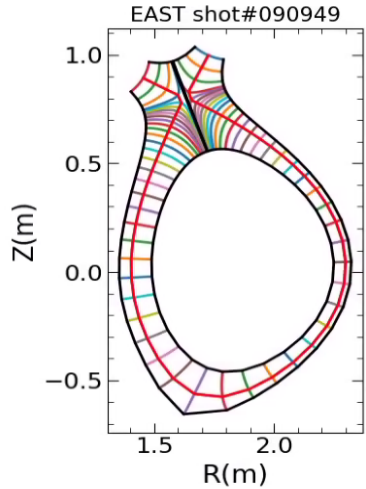

Fig 1. Magnetic geometry and grid for BOUT++ simulation.

The detailed main plasma parameters are shown in table 1 with shot numbers #90949. According to the existing tokamak experiments, this parameter space overlaps with that of the grassy-ELM regime. Key parameters of this discharge are summarized in table 1. The duration of this discharge is last for 60s with high grassy ELM frequency ~ 1kHz, high $\beta_p$~2, high $q_{95}$~7 and improved confinement performance $H_{98y2}$=1.3. This shot is RF-only fully non-inductive with high density and collisionality. The radial equilibrium profiles of pressure and current density for this shot are shown in Fig. 2(a). The initial plasma profiles inside the separatrix of the shot used in these simulations are taken from fits of a modified tanh function to experimental data, mapped onto a radial coordinate of normalized poloidal magnetic flux. For purposes of studying pedestal turbulence, the initial profiles for ψ > 1.0 extend smoothly into the SOL region with small constant gradient. The flux-limited parallel thermal transport is implemented in the simulations. The initial plasma profiles of density and temperature in the simulations are shown in Fig. 2(b). Here the ion temperature is assumed equaled to electron temperature.

In order to understand the ideal MHD instability for grassy ELM, linear simulations are performed with both ELITE[18] and BOUT++ two-fluid three-field turbulence code[19]. The detail results are shown in next section 3.1. To calculate the edge plasma transport and divertor heat flux width, BOUT++ two-fluid six-field turbulence code is used to evolve the perturbations of vorticity $\varpi$, ion density $n_i$, ion and electron temperature $T_i$ and $T_e$, ion parallel velocity $V_{\parallel i}$ and parallel magnetic vector potential $A_\parallel$, while keeping the equilibrium part of the variables unchanged. However, the fluctuation part of profiles evolves self-consistently which modifies the equilibrium part. In the simulations, sheath boundary conditions are imposed on the divertor targets. Neumann boundary condition is applied on inner radial boundary while Dirichlet boundary condition for outer radial boundary. For the core region, twist-shift periodic boundary condition is set in y direction and periodic boundary condition is used in toroidal direction. More detailed settings can be found in previous papers[20].

Table 1 The most important parameters feature EAST shot #90949.

| shot | $f_{ELM}$ | $H_{98y2}$ | $\beta_p$ | $\beta_N$ | $f_{BS}$ | $q_{95}$ | $n_e/n_{GW}$ | $n_{e,sep}/n_{e,ped}$ |
|---|---|---|---|---|---|---|---|---|
| #90949 | ~1kHz | 1.3 | 2.0 | 1.6 | 45% | 7 | 0.8 | 0.6 |

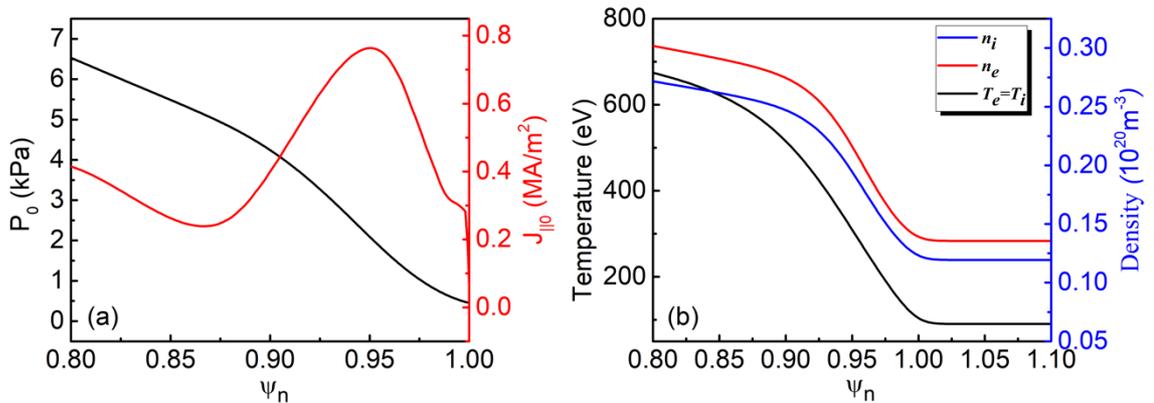

Fig 2. Plasma profiles for BOUT++ simulation: (a) the pressure (black) and current density (red) profiles; (b) ion density (blue), electron density (red) and temperature (black) profiles with $T_i = T_e$.

## 3. Characteristics of Grassy ELM for EAST 90949 shot

### 3.1 Linear MHD stability analysis

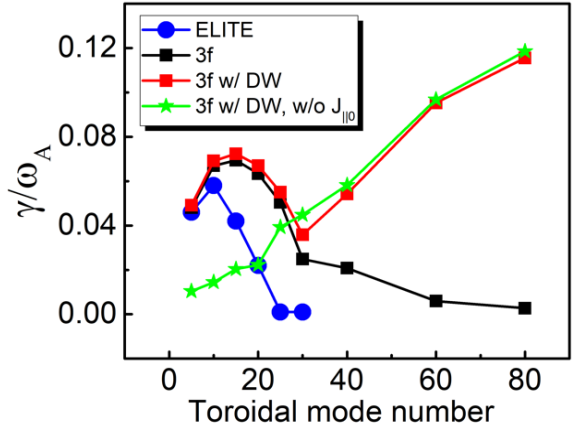

Fig 3. Normalized linear growth rate vs toroidal mode number for various physics models. The blue curve is from ELITE code and the black, red and green curves from BOUT++. The black curve is for ideal MHD, while the red curve is for ideal MHD with drift-Alfvén instability and the green curve is for drift-Alfvén instability without peeling instability.

In order to determine the dominated linear instability drive for the grassy ELMs, we first focus on linear simulations by turning off the nonlinear terms in 3-field turbulence code. The linear simulations start from small initial perturbations with random phases. Fig. 3 shows the toroidal mode number spectrum of the linear growth rate. The horizontal axis is the toroidal mode number and the vertical axis is the mode growth rate normalized by the Alfven frequency at the magnetic axis, $\omega_A = \frac{B_0}{R_0\sqrt{\mu_0 n m_i}}$. BOUT++ linear simulations with ideal MHD instability show that perturbations first grow up around the location of the peak gradient of pedestal pressure at the outer-mid plane and the most unstable toroidal mode numbers in the range of n=10-15 with characteristics of peeling-ballooning modes as shown by the black curve in Fig.3, which shows a similar trend with ELITE as shown by blue curve. The linear simulations of BOUT++ 3-field two-fluid turbulence code further find that drift-Alfvén instabilities yield significant contribution to the high-n modes with n=40-80 as shown by the red and green curves in Fig. 3. Here the red curve shows the modes spectrum with both ideal peeling-ballooning and drift-Alfvén instabilities, while the green curve without peeling mode. By comparing the red and green curves with the black curve, we can find that, with the drift-Alfvén instability, the linear growth rate for high-n modes is larger than that either with or without peeling mode. Because the peeling instability dominates at low-n modes and the drift-Alfvén instability dominates high-n modes for linear simulations, their impact on the ELM nonlinear dynamics is significant different. More details about the nonlinear analysis will be presented in section 3.3.

**3.2 Characteristics of ELM nonlinear dynamics**

BOUT++ two-fluid six-field turbulence nonlinear simulations are conducted to capture the physics of the grassy ELM dynamics and its impact on the divertor heat flux width. The same equilibrium profiles are used in the nonlinear simulations as shown in Fig. 2. The radial simulation domain covers from normalized poloidal flux $\psi_N = 0.8$ to $\psi_N = 1.1$, including the pedestal, the SOL, and private flux regions. The nonlinear simulations start from small initial perturbations with random phases.

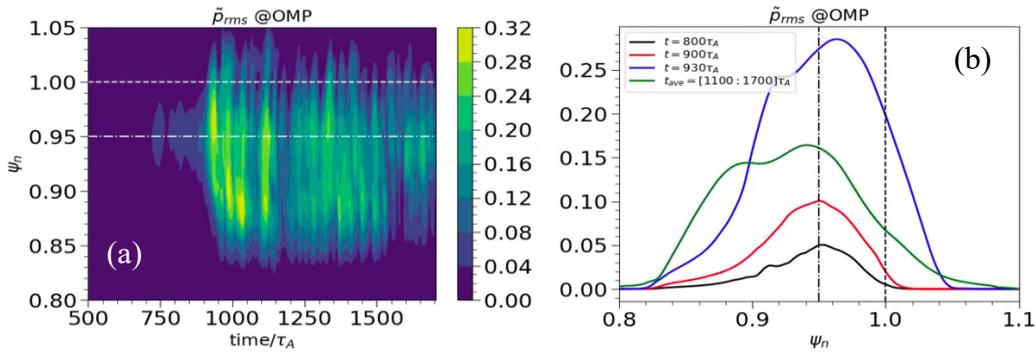

Fig 4. (a) Spatial-temporal evolution of RMS value of pressure fluctuation vs radius and time at the outer midplane (OMP). The white dashed line indicates the location of separatrix and the dash-dot line indicates the peak gradient location of the pressure; (b) Radial mode structures of pressure fluctuation at OMP at different time slices.

The spatial-temporal evolution of the pressure fluctuation further illustrates the generation of the turbulence. In Fig. 4(a), the root-mean-square (RMS) value of pressure fluctuation is obtained at the outer midplane (OMP) and normalized by equilibrium pressure at the pedestal top (4470.7 Pa). The linear phase lasts for about ~$450\tau_A$, the fluctuation saturates at a low level without an ELM triged. While at ~$750\tau_A$, low-n mode grows up at outer midplane and the fluctuation level increases. Finally, the fluctuations saturate at a relative high level, leading to collapse of the pedestal pressure profile and an ELM event. In later nonlinear stage, the fluctuation level saturates at around 16% near the location of the peak pedestal gradient. The radial mode structures of pressure fluctuation at outer midplane in the linear and nonlinear stage are shown in Fig. 4 (b). Before the ELM crash, the mode grows up at the peak pressure gradient location and the perturbation is localized within the pedestal region as shown by the black and red curves in Fig.4(b). While in the later nonlinear stage, the perturbation radially spreads on both sides of the pedestal region, reaching into the inner edge region and the SOL, indicating the collapse of pressure pedestal as shown by the blue and green curves in Fig. 4(b). The time evolution of radial pressure profile demonstrates the profile flattening processes during the ELM bursting activity. The pressure profile keeps dropping inside the separatrix while rising in the SOL, indicating that energy is continuously transported from the edge region to the SOL. In the SOL, the energy is transported along the magnetic field line and is finally deposited onto the divertor targets.

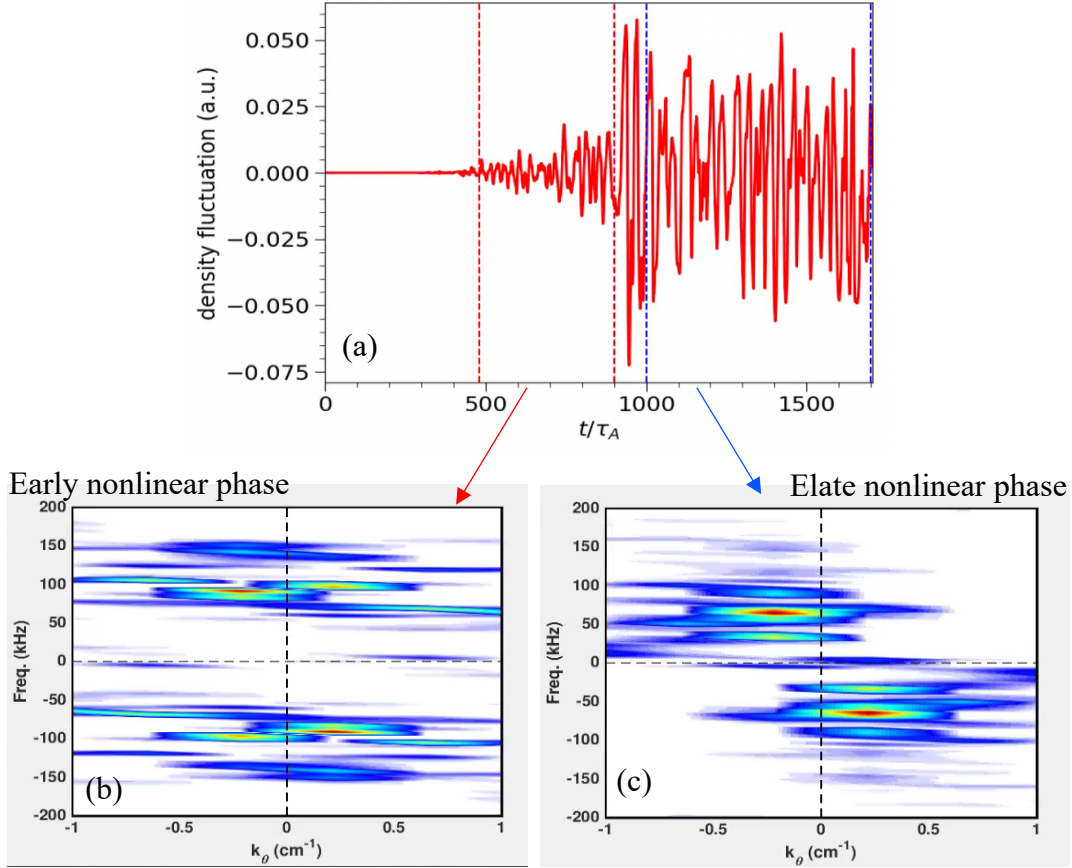

Fig. 5 (a) The time evolution of density fluctuation; the wavenumber spectrum for density fluctuation at (b) early nonlinear phase and (c) late nonlinear phase. The data is taken at the peak pressure gradient position at outer midplane.

By closely examining the time history of density fluctuation, two different nonlinear saturation stages are found from this grassy ELM simulation, as shown in Fig. 5(a): (1) the early nonlinear stage from $t = 480\tau_A$ to $t = 900\tau_A$ as shown between the red-dashed lines; (2) the late nonlinear stage after $t = 1000\tau_A$, as shown between the blue-dashed lines. In order to identify the characteristics of two different nonlinear stages, the spectrum of density fluctuation is calculated from the two-point correlation function with two different poloidal grid points at OMP. Figure 5(b) and (c) show the contour plot of density fluctuation vs wave number and frequency in different nonlinear stages. In the early nonlinear stage, the k-f spectrum is calculated with the time range $480\tau_A - 900\tau_A$. Density fluctuation saturates in a low level with high frequency modes, corresponding to drift-Alfvén modes. The mode frequency is ~80-

100 kHz and the poloidal wave number $k_\theta$ is about 0.25 cm as shown in Fig. 5(b). At the late nonlinear stage, density fluctuation saturates in a high level with several mode frequencies (~ 20, 60 and 100kHz) and a similar range of poloidal wave number $k_\theta$, as shown in Fig. 5(c). These modes correspond to peeling-ballooning modes, generate large radial transport, and thus trigger an ELM. Because low-n peeling-ballooning modes have lower linear growth rates than drift-Alfvén instabilities, they grow up late in the initial value simulation code.

**3.3 Impact of P-B mode and drift-Alfvén instabilities on ELMs**

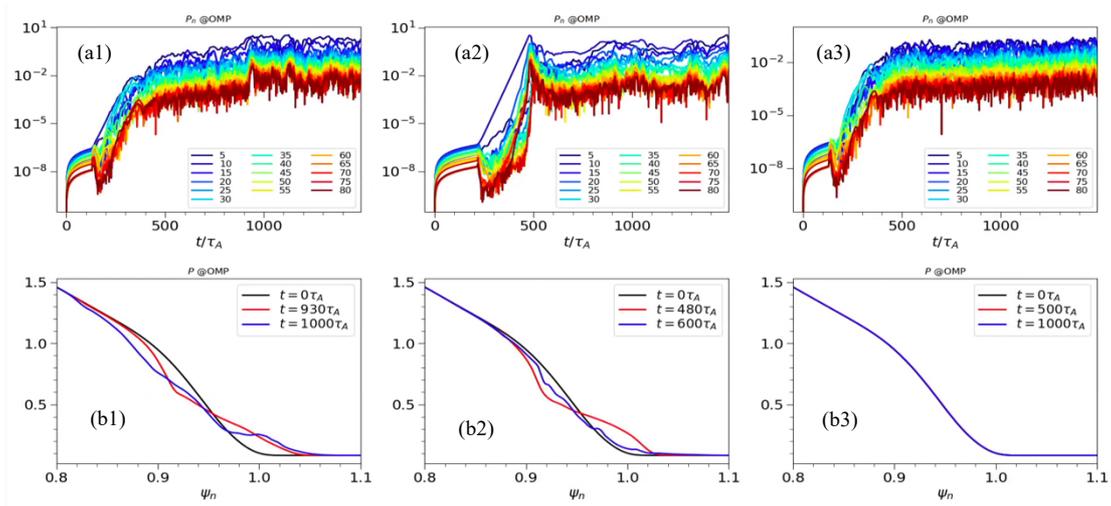

Fig 6. (a) Time evolution of pressure fluctuation for different toroidal mode numbers and (b) The radial pressure profiles normalized by equilibrium pressure at the pedestal top (4470.7 Pa) at different times. The three columns are for three different linear instability drives: (a1, b1) with both peeling and drift-Alfvén drives; (a2, b2) with peeling drive only without drift-Alfvén drive; (a3, b3) with drift-Alfvén drive only without peeling drive.

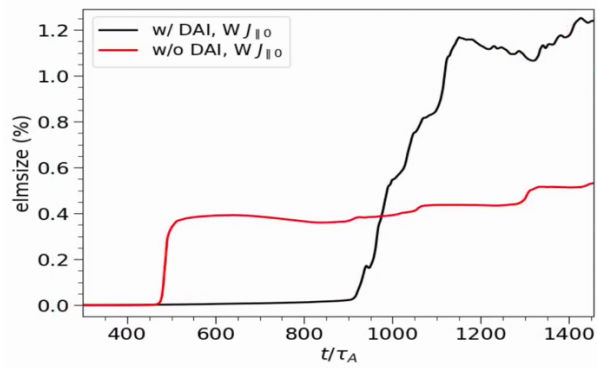

Fig. 7 Time evolution of 3D relative ELM size with different linear instability drive. The black curve is the case with both peeling and drift-Alfvén drives while the red curve is the case with peeling drive without drift-Alfvén drive.

In order to understand the relative role of peeling-ballooning instability and drift-Alfvén instability in the ELM dynamics, three nonlinear simulations are performed with turning on or off the drift-Alfvén (or "eHall" in the code input option) and peeling (or "current" in the code input option) terms in the equations. Fig 6 (a1) and (b1) show the simulation results with both peeling and drift-Alfvén instabilities; (a2) and (b2) show the simulation results without drift-Alfvén instability, while (a3) and (b3) show the simulation results without peeling instability. Without peeling drive in Fig. 6 (a3) and (b3), the fluctuation amplitude drops by an order of magnitude and no ELM crash occurs in comparison with Fig. 6 (a1) and (b1). The pressure profile remains unchanged throughout simulation as shown in Fig.6 (b3); while the pressure profile collapses with peeling drive as shown by Fig 6 (b1) and (b2). Therefore, from these three nonlinear simulations, we can find that the ELM crash is trigged by peeling mode for this grassy ELM shot even though the drift-Alfvén instabilities dominate the linear growth phase in the initial value simulation code. From the simulation we further find that the drift-Alfvén instability will delay the onset of the ELM and enhance the turbulence transport by comparing the first two columns in Fig.6. The fluctuation extends inward beyond peak gradient region with drift-Alfvén instability in Fig. 6 (b1) in comparison with that without it in Fig. 6 (b2) even though their fluctuations saturate in the same level at the later nonlinear stage. Without the drift-Alfvén instability, the ELM can be easily triggered around 480 $\tau_A$ and the relative ELM size, or the energy loss fraction, will be saturated at 0.5% as shown by the red curve in Fig.7. While with the drift-Alfvén instability, the fluctuation saturates at a low level first and then grows up again, and finally trigger an ELM crash around $840\tau_A$. The onset of ELM is delayed due to the drift-Alfvén turbulence, which acts as a damping effect to reduce the linear growth rate of peeling modes. The ELM size is saturated at a high level ~1.2% with a strong turbulence transport enhanced by drift-Alfvén turbulence as shown by the black curve in Fig 7. From the nonlinear simulations, we also find the peeling modes and drift-Alfvén instabilities grow up at different poloidal locations. Fig. 8 shows the mode structure at different stages. Fig 8 (a) shows the mode structure at linear stage $t = 250\tau_A$ and (b) shows the mode structure before ELM crash $t = 840\tau_A$. Drift-Alfvén instability is generated and grows up first with high growth rate in the high-field side (HFS) with high-n mode numbers as shown by Fig 8(a) and the mode structure is consistent with the case of without peeling drive. There is no ELM crash when it gets into the nonlinear Drift-Alfvén stage with a lower lever fluctuation. While in the late nonlinear Drift-Alfvén stage, the peeling mode grows large in the low-field side (LFS) with low-n mode. The ELM is triggered and the dominated mode structure is comparable with the case without drift-Alfvén instability.

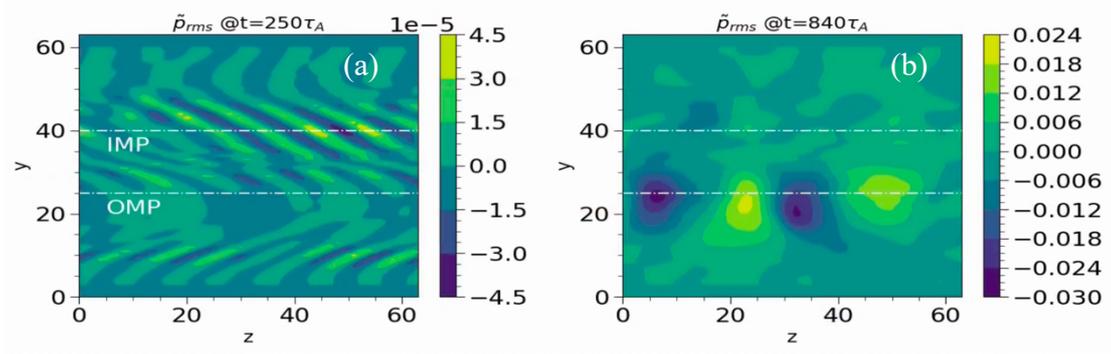

Fig 8. Contour plot of pressure fluctuation vs poloidal and binormal directions at different stages for the case with both peeling and drift-Alfvén instabilities, (a) at the linear stage $t = 250\tau_A$ (b) late nonlinear Drift-Alfvén stage before an ELM crash at $t = 840\tau_A$.

## 4. Divertor heat flux width

In order to investigate the impact of the grassy ELM on the divertor heat load, the BOUT++ simulates the evolution of the heat flux during one burst of ELM event. We track the evolution of heat load on the target during an ELM event and both the particle flux width and heat flux width are calculated in the saturation phase on the outer divertor target. Direct comparisons of simulated particle flux with experimental data are shown in Fig.9. The open circles are calculated from BOUT++ simulations and the blue curve is a fit to the profile using the Eich fitting formula[21,22]. The particle flux width calculated by BOUT++ is ~10.6mm which is comparable to heat flux width. The six-pointed purple stars and black stars are the experimental data points for grassy ELM discharges, measured by Langmuir probes. The magnitude of the particle flux calculated by BOUT++ is comparable with experimental measurements while the width is around 1-3 times smaller than the experimental results. One possible reason for the difference is caused by RF heating, which can change the magnetic topology. Magnetic topology changes induced by lower hybrid waves (LHW) can lead to splitting of the strike point and result in a dramatic broadening of the heat flux width[12,23]. The LHW effect is not considered in our model. In addition, it worth to mention that the experiment data is limited for this grassy ELM discharge. More experimental data are needed to establish a statistical average for further comparison.

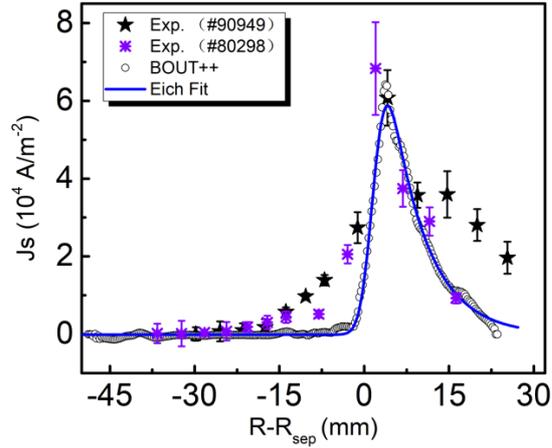

Fig. 9 Particle flux at outer divertor target. The open circles are calculated from BOUT++ simulations and the blue curve is a fit to the profile using the Eich fitting formula[21]. The six-pointed purple stars and black stars are the experimental measurements by Langmuir probes for two different discharges.

Table 2. The SOL width based on different physics models.

| Shot | East #90949 |
|---|---|
| $v_{e,SOL}^*$ | 6.87 |
| Eich: $\lambda_q$(mm) | 5.5 |
| HD/GHD: $\lambda_q$(mm) | 5.7/6.3 |
| $\lambda_q^{eich,tur}$(mm) | 10.01 |
| BOUT++ transport: $\lambda_q^{transport}$(mm) | 13.64 |
| BOUT++ turbulence: $\lambda_q^{turb}$(mm) | 10.67 |

The heat flux width is calculated with different physics models to compare the heat flux width broadening mechanisms. To calculate the divertor heat flux width from simulations, we apply Eich's fitting formula to the parallel electron heat flux profile in steady-state phase of nonlinear simulations. Divertor heat flux width calculated based on different models are shown in table 2. From the empirical Eich scaling[21,22] and HD model[24], the heat flux width is 5.5 mm and 5.7 mm respectively. With the effect of collisionality on parallel transport, the heat flux width will be broadened by ~11% calculated by the GHD model[25,26] while with the effect of collisionality on radial turbulent transport the width is enhanced by a factor of 2[27]. Here the collisionality is calculated by $v^*_{e,SOL} = \frac{\pi \hat{q}_{cyl} R}{1.03 \cdot 10^{16}} \frac{n_e}{T_e^2} Z_{eff}$. Eich turbulence regression is based on the reference[27]:

$$\frac{\lambda_p}{\rho_{s,pol}} = \left(1 + (3.6 \pm 0.19)\alpha_t^{1.9 \pm 0.14}\right) \cdot (1.2 \pm 0.05); \qquad \alpha_t \simeq \frac{1}{100} \cdot \hat{q}_{cyl} v_e^*.$$

Both BOUT++ transport and turbulence simulations show the heat flux width for grassy ELM is broadened 2-3 times due to the large radial turbulent transport. From the turbulence simulations, we find the effective radial turbulence transport coefficient $\chi_{e,tur} = 3.1$ m$^2$/s, which is larger than the critical value $\chi_{\perp,critical} = 1.02$ m$^2$/s calculated from the formula in the reference[10] for the transition from drift-dominated regime to turbulent dominated regime. Therefore, the heat flux width will not obey the HD-based empirical Eich scaling. It worth to mention that, even though the BOUT++ simulations and Eich turbulence model show the similar broadening (~2-3 times) for the heat flux width, the underlying physics are different. The Eich turbulence model is based on the resistive interchange & drift-Alfvén turbulence. While for the BOUT++ turbulence simulations, the width is broadened due to the large turbulence driven by ideal peeling-ballooning modes with significant contribution from drift-Alfvén turbulence. Fig.10 shows the time evolution of power deposited on the divertor targe with and without drift-Alfvén instability. For the case without drift-Alfvén instability, the power loading increases first after the ELM crashes, and then decays to ~1/e times of peak heat flux. While with drift-Alfvén turbulence, more power is transported from pedestal to SOL, which is then transported along the magnetic field line onto the divertor targets. The power loading on the divertor increases and then is saturated due to the continuous drift-Alfvén turbulence transport.

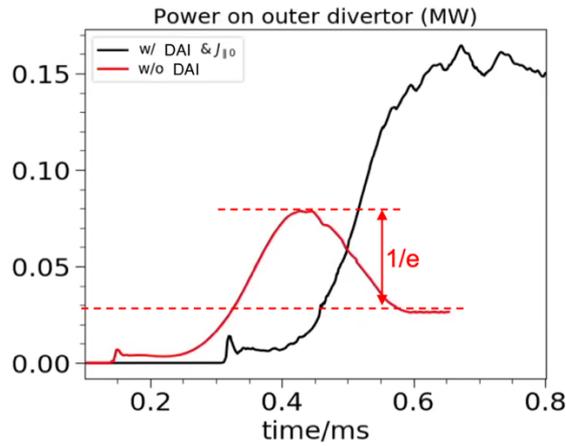

Fig. 10 Time evolution of power deposited on the target. The black curve is the case with both Peeling and drift-Alfvén linear instability drives while the red curve is the case with peeling drive only without drift-Alfvén instability drive.

## 5. Summary

BOUT++ turbulence simulations are conducted to uncover the fluctuation characteristics of the grassy-ELMs and its impact on the divertor heat flux width for EAST High-$\boldsymbol{\beta_p}$ long-pulse grassy-ELM discharges. BOUT++ linear simulations show that the unstable modes cover a range from low-n (n=10~15) with characteristics of peeling-ballooning modes (P-B) to high-n (n=40~80) modes driven by drift-Alfvén instabilities. P-B modes are generated in the low-field side (LFS) with low-n modes while Drift-Alfvén instabilities are generated in the high-field side (HFS) with high-n modes. Even though the drift-Alfvén instabilities dominate the linear growth phase with a wide n-spectrum and the fluctuation peaks on high-field side, nonlinear simulations show that the ELM crash is trigged by peeling mode on low-field side and fluctuation is radially localized near the bottom of pedestal. However, the drift-Alfvén turbulence delays the onset of the ELM crash, the energy loss increases with drift-Alfvén turbulence, and the fluctuation extends to peak gradient region. The saturated ELM size is 0.5% without drift-Alfvén instability while it becomes larger with drift-Alfvén instability due to the enhancement of radial turbulence transport after the ELM crash. Simulations further show that if the peeling drive is removed, the fluctuation amplitude drops by an order of magnitude and the ELM crashes disappear.

BOUT++ turbulence nonlinear simulations show that the turbulence is generated from the peak gradient of pedestal. The pedestal plasma particles and energy are transported across the separatrix into the SOL, which are then transported along the magnetic field lines, and finally deposited on the divertor plates. The simulated particle flux width is comparable with heat flux width. The magnitude of the particle flux calculated by BOUT++ is comparable with experimental measurements while the width is ~1-3 times smaller than that of the experimental measurements. The divertor heat flux width given by both BOUT++ transport and turbulence simulations are about 2~3 times larger than the estimates based on the HD model and the Eich's ITPA multi-tokamak scaling due to strong fluctuations in grassy ELM regimes. For the case without drift-Alfvén instability, the power loading increases first after ELM crash then decays to ~1/e times of peak heat flux, indicating that the power loading possibly fluctuates by 50% during the grassy ELMs. This level of fluctuation is much smaller than that of type-I ELMs, which is a pulsed heat load with an amplitude variation by 10 to 100 times. While with drift-Alfvén instability, more power is slowly transported from pedestal to the SOL and the power loading on the divertor keeps more or less steady, indicating that the power loading is continuous.

**Acknowledgement**

This work was supported by the National Key R&D Program of China Nos. 2017YFE0301206, 2017YFE0300402 and 2017YFE0301100 and National Natural Science Foundation of China under Grant No. 11675037. This work was also performed under the U.S. Department of Energy by Lawrence Livermore National Laboratory under Contract No. DE-AC52-07NA27344.